\documentclass[12pt,a4paper]{article}
\usepackage{latexsym,epsfig,graphicx,a4wide,amssymb,amsmath,euscript}

\def\be{\begin{equation}}
\def\ee{\end{equation}}
\def\bea{\begin{eqnarray}}
\def\eea{\end{eqnarray}}
\DeclareMathOperator{\arctanh}{arctanh}
\def\gen{\mathrm{g}}
\def\Pcmn{\mathcal{P}_{l}^{mn}(\cosh\theta)}
\def\Pc{\mathcal{P}^{m0}_{l}}

\def\Pcksi{\mathcal{P}^{m0}_{-\frac{1}{2}+i\xi}}

\def\Ycksi{\mathcal{Y}_{\xi}^{m}}
\def\Ycksip{\mathcal{Y}_{\xi'}^{m'}}
\def\Tcov{\tilde{T}}

\numberwithin{equation}{section}

\catcode`\@=11

\def\section{\@startsection{section}{1}{\z@}{3.5ex plus 1ex minus
   .2ex}{2.3ex plus .2ex}{\large\bf}}

\def\ps@headings{\def\@oddfoot{}\def\@evenfoot{}
\def\@oddhead{\hbox{}\hfill
        \makebox[.5\textwidth]{\raggedright\ignorespaces --\thepage{}--
        \hfill }}
\def\@evenhead{\@oddhead}
\def\subsectionmark##1{\markboth{##1}{}}
}

\ps@headings

\catcode`\@=12

\begin{document}

\begin{titlepage}
\rightline{December 2008}

\begin{centering}
\vspace{1.4cm}


{\Large{\bf Hawking Radiation via Gravitational Anomalies in
Non-spherical Topologies }\\}

\vspace{2cm}


{\bf Eleftherios Papantonopoulos$^{*}$ and Petros Skamagoulis}$^{**}$\\
\vspace{.2in}
 Department of Physics, National Technical University of Athens,\\
 Zografou Campus GR 157 73, Athens, Greece. \\
\vspace{3mm}

\end{centering}
\vspace{2.2cm}


\begin{abstract}

We study the method of calculating the Hawking radiation via
gravitational anomalies in gravitational backgrounds of constant
negative curvature. We  apply the method to topological black
holes and also to topological black holes conformally coupled to a
scalar field.

\end{abstract}

\vspace{6.0cm}
\begin{flushleft}
$^{*}~~$ Electronic address: lpapa@central.ntua.gr \\
$^{**}~$ Electronic address: pskam@central.ntua.gr \\
\end{flushleft}

\end{titlepage}


\section{Introduction}

Hawking radiation is an important quantum effect in black hole
physics. It arises for quantum fields in a background spacetime with
an event horizon. Apart from Hawking's original
derivation~\cite{Hawking:1974sw,Hawking:1974rv}, which calculates
the Bogoliubov coefficients between in and out states for a body
collapsing to form a black hole, there are also other
approaches~\cite{Gibbons:1976ue},~\cite{Parikh:1999mf}. One of the
most interesting proposals was put forward many years ago by
Christensen and Fulling~\cite{Christensen:1977jc}, who showed that
Hawking radiation can be derived from the trace anomaly in the
energy-momentum tensor of quantum fields in a Schwarzschild black
hole background.

The idea of Christensen  and Fulling \cite{Christensen:1977jc} was
to relate  an anomaly in conformal symmetry with the energy-momentum
tensors of quantum fields in a black hole background.  This relation
manifests itself as a contribution of the anomaly to the trace
$T^\alpha_\alpha$ of the energy-momentum tensor in a theory where it
vanishes classically. Requiring finiteness of the energy-momentum
tensor of massless fields as seen by a freely falling observer at
the horizon in (1+1)-dimensional Schwarzschild background metric and
using the anomalous trace equation everywhere, one finds an outgoing
flux which is in quantitative agreement with Hawking's result.

The validity of this result is subjected to some limitations. The
method has been applied  to conformal field theories in (1+1)
dimensions. Also, the assumption in~\cite{Christensen:1977jc} of
massless scalar fields was essential to relate fluxes at the horizon
to Hawking radiation. The requirement of massless scalar fields was
addressed later in~\cite{ChiouLahanas:1995ec}. It was considered a
massive tachyon field in the  background of a dilatonic
(1+1)-dimensional black hole~\cite{Diamandis:1994uc}. It was found
that the contribution of the tachyon field to the Hawking radiation
is due to its coupling to the dilaton field, and the Hawking rate
due to the tachyon field is enhanced comparable to conformal matter.

Quite recently, Robinson and Wilczek~\cite{Robinson:2005pd} followed
by Iso, Umetsu and Wilczek~\cite{Iso:2006wa} proposed a new method
to calculate the Hawking radiation. Their basic idea is to identify
outgoing modes of some matter distribution near the horizon as right
moving modes while ingoing modes as left moving modes, in the Unruh
vacuum~\cite{Unruh:1976db}. Then, because all the ingoing modes can
not  classically affect physics outside the horizon, integrating the
other modes they obtain an effective chiral action in the exterior
region which is anomalous under gauge and general coordinate
transformations. However, the underlying theory is invariant under
these symmetries and these anomalies must be cancelled by quantum
effects of the classically irrelevant ingoing modes.  They have
proved that the condition for anomaly cancellation at the horizon
determines the Hawking flux of the charge and energy-momentum. The
flux is universally determined only by the value of anomalies at the
horizon.

The crucial point in the Robinson-Wilczek method and its
generalization to include charge, is to reduce an initially
high-dimensional theory to two dimensions, in the vicinity of the
horizon, which is a necessary step in order to be able to identify
the chiral modes. This is achieved by considering a matter source,
just outside the horizon of a static spherically symmetric black
hole, parametrized by a scalar field minimally coupled to this
background. Performing a partial wave decomposition of the scalar
field in terms of the wavefunctions of the classical wave equation,
they find that the effective radial potentials for partial wave
modes of the scalar field vanish exponentially fast near the
horizon. Thus, physics near the horizon can be described using an
infinite collection of massless (1+1)-dimensional scalar fields,
each propagating in a (1+1)-dimensional spacetime with a metric
given by the $(t,r)$ section of the original high-dimensional
metric, where $t$ and $r$ are the time and radial coordinate
respectively.

The method was further extended to include
rotations~\cite{Iso:2006ut, Murata:2006pt}. It was shown that the
reduction procedure goes through and observing that an angular
isometry generates an effective $U(1)$ gauge field in the
(1+1)-dimensional theory, with the azimuthal quantum number $m$
serving as the charge of each partial wave, the known results were
obtained with angular momentum acting like a chemical potential for
the effective charge.

In this work we will study the Hawking effect via the gravitational
anomalies method  in a gravitational background of constant negative
curvature. At first, we will carry out the dimensional reduction
procedure of an action given by a  scalar field minimally coupled to
gravity in the background of a (3+1)-dimensional topological black
hole (TBH), in order to show that near the horizon the theory is
reduced to an effective theory of an infinite collection of
(1+1)-dimensional scalar fields in a (1+1)-dimensional background.
Identifying the chiral modes we will finally show that the flux
necessary to cancel the gravitational anomalies is identified with
the Hawking flux.

We will also apply the method to a topological black hole coupled to
a scalar field. Providing that asymptotically the space is AdS,
these black hole solutions are stable, they
 satisfy the
Breitenlohner-Freedman bound~\cite{Breitenlohner:1982jf} and the
scalar field is regular at the horizon~\cite{Martinez:2004nb}. In
this context, we will discuss the applicability of the method in the
case of a scalar field backreacting on the gravitational background.
As a first step we will consider the case where the scalar field,
which generates the Hawking flux, is non-minimally coupled to
gravity.

The paper is organized as follows. In section 2 we review the basic
properties of TBHs and we perform a mode analysis of a scalar field
in the background of a TBH. In section 3 we describe the reduction
procedure to two dimensions for a TBH of genus $\gen=2$. In section
4 we derive the Hawking radiation of a TBH of genus $\gen=2$ and in
section 5 we carry out the same calculation for a TBH coupled to a
scalar field. In section 6 we investigate whether a scalar field
non-minimally coupled to a black hole background has any effect on
the Robinson-Wilczek method. Finally, we summarize in the last
section.


\section{Mode Analysis of the Wave Equation of a Scalar Field in the Background of a TBH}

We consider the bulk action
\begin{equation}
I=\frac{1}{16\pi G}\int d^{d}x \sqrt{-g}
\biggl[\mathrm{R}+\frac{(d-1)(d-2)}{l^{2}}\biggr]~,
\end{equation}
in asymptotically AdS$_{d}$ where $G$ is the Newton's constant,
$\mathrm{R}$ is the Ricci scalar and $l$ is the AdS radius. The
presence of a negative cosmological constant $(\Lambda =-
\frac{(d-1)(d-2)}{2l^{2}})$ allows the existence of black holes with
topology $\mathbb{R}^{2}\times\Sigma$, where $\Sigma$ is a
$(d-2)$-dimensional manifold of constant negative curvature. These
black holes are known as topological black holes~\cite{topological,
Vanzo:1997gw}. The simplest solution of this kind in four dimensions
reads
\begin{equation}\label{linel}
ds^{2}=-f(r)dt^{2}+\frac{1}{f(r)}dr^{2}+r^{2}d\sigma ^{2}\quad
\textrm{,\quad $f(r)=r^{2}-1-\frac{2\mu}{r}$}~,
\end{equation}
where we employed units in which the AdS radius is $l=1$, $\mu$ is a
constant which is proportional to the mass and is bounded from below
$\mu\geq-\frac{1}{3\sqrt{3}}$ and $d\sigma^{2}$ is the line element
of the two-dimensional manifold $\Sigma$, which is locally
isomorphic to the hyperbolic manifold $H^{2}$ and of the form
\begin{equation}
\Sigma=H^{2}/\Gamma \quad \textrm{,\quad  $\Gamma\subset O(2,1)$}~,
\end{equation}
where $\Gamma$ is a freely acting discrete subgroup (i.e. without
fixed points) of isometries. The line element $d\sigma^{2}$ of
$\Sigma$ is
\begin{equation}
d\sigma^{2}=d\theta^{2}+\sinh^{2}\theta d\varphi{^2}~,
\end{equation}
with $\theta\ge0$ and $0\le\varphi<2\pi$ being the coordinates of
the hyperbolic space $H^{2}$ or pseudosphere, which is a non-compact
two-dimensional space of constant negative curvature. This space
becomes a compact space of constant negative curvature with genus
$\gen\ge2$ by identifying, according to the connection rules of the
discrete subgroup $\Gamma$, the opposite edges of a $4\gen$-sided
polygon whose sides are geodesics and is centered at the origin
$\theta=\varphi=0$ of the pseudosphere~\cite{topological,
Vanzo:1997gw, Balazs:1986uj}. An octagon is the simplest such
polygon, yielding a compact surface of genus $\gen=2$ under these
identifications. Thus, the two-dimensional manifold $\Sigma$ is a
compact Riemann 2-surface of genus $\mathrm{g}\geq2$. Further
details on this kind of compactification scheme can be found
in~\cite{Balazs:1986uj,Sausset:2007hh}. The configuration
(\ref{linel}) is an asymptotically locally AdS spacetime. The
horizon structure of (\ref{linel}) is determined by the roots of the
metric function $f(r)$ that is
\begin{equation}
f(r)=r^{2}-1-\frac{2\mu}{r}=0~.
\end{equation}
For $-\frac{1}{3\sqrt{3}}<\mu<0$, this equation has two distinct
non-degenerate solutions, corresponding to an inner and an outer
horizon $r_{-}$ and $r_{+}$ respectively. For $\mu\geq0$, $f(r)$ has
just one non-degenerate root and so the black hole (\ref{linel}) has
one horizon $r_{h}$. The horizons for both cases of $\mu$ have the
non-trivial topology of the manifold $\Sigma$. We note that for
$\mu=-\frac{1}{3\sqrt{3}}$, $f(r)$ has a degenerate root, but this
horizon does not have an interpretation as black hole
horizon~\cite{topological}.

We will examine the eigenmodes of the classical wave equation of a
scalar field $\Phi$ of mass $m_{\Phi}$, in the background of the
topological black hole (\ref{linel}) and perform a partial wave
decomposition of the wavefunctions. The classical wave equation in
the background of (\ref{linel}), without any identifications of the
pseudosphere (i.e. $H^{2}$), is
\begin{equation}
\nabla^{2}\Phi=m_{\Phi}^{2}\Phi~,
\end{equation}
where $\nabla^{2}$ is the Laplace-Beltrami operator defined by
\begin{equation}
\nabla^{2}\equiv\frac{1}{\sqrt{-g}}\partial_\mu\bigl(\sqrt{-g}g^{\mu\nu}\partial_\nu\bigr)~,
\end{equation}
and hence the wave equation is
\begin{equation}\label{wveq}
\biggr[-\frac{1}{f}\partial_{t}^{2}+\frac{1}{r^{2}}\partial_{r}\big(r^{2}f\partial_{r}\big)
+\frac{1}{r^{2}\sinh\theta}\partial_{\theta}\big(\sinh\theta\partial_{\theta}\big)+
\frac{1}{r^{2}\sinh^{2}\theta}\partial_{\varphi}^{2}\biggl]\Phi=m_{\Phi}^{2}\Phi~.
\end{equation}
We factorize out the angular and radial dependence of the field as
\begin{equation}
\Phi(t,r,\theta,\varphi)=\frac{R(t,r)}{r}\mathrm{Y}(\theta,\varphi)~.
\end{equation}
With this factorization and using separation of variables we get
two differential equations. The angular wave equation is
\begin{equation}\label{angwve}
-\bigg[\frac{1}{\sinh\theta}\partial_{\theta}\big(\sinh\theta\partial_{\theta}\big)+
\frac{1}{\sinh^{2}\theta}\partial_{\varphi}^{2}\bigg]\mathrm{Y}(\theta,\varphi)=\lambda
\mathrm{Y}(\theta,\varphi)~,
\end{equation}
while the radial wave equation is
\begin{equation}
\partial_{t}^{2}R(t,r)-f\bigg[(\partial_{r}f)\partial_{r}-\frac{\partial_{r}f}{r}+f\partial_{r}^{2}
-m_{\Phi}^{2}-\frac{\lambda}{r^2}\bigg]R(t,r)=0~,
\end{equation}
where $\lambda$ is a separation constant. The angular wave equation
has the solution~\cite{Balazs:1986uj}
\begin{equation}\label{angsol}
\mathrm{Y}_{l}^{m}(\theta,\varphi)=P_{l}^{m}(\cosh\theta)e^{im\varphi}
=P_{-\frac{1}{2}\pm i\xi}^{m}(\cosh\theta)e^{im\varphi}~,
\end{equation}
where $P_{l}^{m}$ are the associated Legendre functions and
\begin{eqnarray}\label{qnumb}
m &=& 0, \pm1, \pm2, \pm3, \ldots \nonumber\\
\lambda &=&-l(l+1)~,\nonumber\\
l &=&-\frac{1}{2}\pm i\xi~, \nonumber\\
\lambda &=&\xi^{2}+\frac{1}{4}~.
\end{eqnarray}
The radial wave equation, after separating off the time dependence
by writing $R(t,r)=R(r)e^{i\omega t}$, becomes
\begin{equation}
\omega^{2}R(r)+f\bigg[(\partial_{r}f)\partial_{r}-\frac{\partial_{r}f}{r}+f\partial_{r}^{2}
-m_{\Phi}^{2}-\frac{\xi^{2}+\frac{1}{4}}{r^2}\bigg]R(r)=0.
\end{equation}
There is no general solution to this equation but we can write it in
a very simple form using the tortoise coordinate $r_{*}$ defined by
\begin{equation}\label{tortoise}
\frac{\partial r_\ast}{\partial r}=\frac{1}{f(r)}~.
\end{equation}
The radial wave equation in terms of the tortoise coordinate $r_{*}$
becomes
\begin{equation}
\bigg[\partial_{r_{*}}^2+\omega^{2}-f\big(r(r_{*})\big)V\big(r(r_{*})\big)\bigg]R(r(r_{*}))=0~,
\end{equation}
with
\begin{equation}
V(r)=\frac{1}{r^{2}}\bigg[r\frac{df(r)}{dr}+\xi^{2}+\frac{1}{4}\bigg]+m_{\Phi}^{2}~,
\end{equation}
which for the $f(r)$ of the background (\ref{linel}) is
\begin{equation}
V(r)=2+\frac{\xi^{2}+\frac{1}{4}}{r^{2}}+\frac{2\mu}{r^{3}}+m_{\Phi}^{2}~.
\end{equation}
In conclusion, the eigenmodes of the classical wave equation in the
background of (\ref{linel}) are
\begin{equation}\label{basis}
\Phi(t,r,\theta,\varphi)=\frac{R_{\xi}(t,r)}{r} P_{-\frac{1}{2}\pm
i\xi}^{m} (\cosh\theta)e^{im\varphi}~.
\end{equation}

Without any identifications of the pseudosphere the spectrum of the
angular wave equation is continuous, thus $\xi$ takes any real value
$\xi\geq0$. Since the two dimensional manifold $\Sigma$ is a
quotient space of the form $H^{2}/\Gamma$ and is a compact space of
constant negative curvature, the spectrum of the angular wave
equation is discretized and thus $\xi$ takes discrete real values
$\xi\geq0$. On the simplest such manifold of constant negative
curvature, that is a compact surface of genus $\gen=2$, the angular
wavefunctions (\ref{angsol}) must satisfy four periodicity
conditions and the compatibility of these four periodicity
conditions is what generates the discrete
spectrum~\cite{Balazs:1986uj}. In general there are no explicit
analytical results in the literature for the angular eigenvalues
$\lambda(l)$ and for the angular eigenfunctions, although some
numerical results exist~\cite{Balazs:1986uj}, so in the next
sections we will elaborate only on the case of $\Sigma$ being a
compact two-dimensional manifold of genus $\gen=2$ with constant
negative curvature.


\section{Dimensional Reduction for a Topological Black Hole}

We consider matter in the background of the topological black hole
(\ref{linel}) of genus $\gen=2$ given by a complex scalar field
$\phi(x)$ with an action of the form
\begin{equation}
S=S_{free}+S_{int}~,
\end{equation}
where $S_{free}$ is the free part of the action
\begin{equation}
S_{free}=-\frac{1}{2}\int d^{4}x\sqrt{-g}\phi^{*}\nabla^{2}\phi~,
\end{equation}
and $S_{int}$ is the part of the action which includes a mass term,
potential terms and interaction terms. We perform a partial wave
decomposition of $\phi$ in terms of the eigenmodes (\ref{basis})
\begin{equation}
\phi(t,r,\theta,\varphi)=\sum_{m=-\infty}^{+\infty}\frac{R_{\xi
m}(t,r)}{r}\Ycksi(\theta,\varphi)~,
\end{equation}
where for convenience we chose a different normalization of the
eigenmodes by defining the functions
$\Ycksi$~\cite{Argyres:1988ga} as
\begin{equation}
\begin{split}
\Ycksi(\theta,\varphi)&\equiv\Bigg(\frac{2\pi}{\xi\tanh(\pi\xi)}\Bigg)^{\frac{1}{2}}\Pcksi(\cosh\theta)e^{im\varphi}\\
&=\Bigg(\frac{2\pi}{\xi\tanh(\pi\xi)}\Bigg)^{\frac{1}{2}}\frac{\Gamma(i\xi+\frac{1}{2})}{\Gamma(i\xi+m+\frac{1}{2})}
P_{-\frac{1}{2}+i\xi}^{m}(\cosh\theta)e^{im\varphi}~,
\end{split}
\end{equation}
with $m=0, \pm1, \pm2, \pm3, \ldots$ and $\xi$ taking discrete real
values $\xi\geq0$. In this definition we have used the functions
$\mathcal{P}_{l}^{mn}$, which form the canonical basis for the
irreducible representations of the group SL$(2,C)$ and can be viewed
as playing the same role for the group SU$(1,1)$ (see Appendix A).
These functions are related to the associated Legendre functions
through
\begin{equation}
\Pc(\cosh\theta)=\frac{\Gamma(l+1)}{\Gamma(l+m+1)}P_{l}^{m}(\cosh\theta)~.
\end{equation}
The functions $\Ycksi$ form a complete set of functions on the
manifold $\Sigma$, they satisfy four periodicity
conditions~\cite{Balazs:1986uj} and their orthogonality condition is
the equation (\ref{orthoApp}) of the Appendix A, that is
\begin{equation}\label{normcond}
\int_{0}^{\infty}d\theta\int_{0}^{2\pi}d\varphi\sinh\theta\Ycksi(\theta,\varphi)
\Big(\Ycksip(\theta,\varphi)\Big)^{*}=\delta_{\xi\xi'}\delta_{mm'}~.
\end{equation}
Furthermore, it is proved in the Appendix A that they satisfy the
equation
\begin{equation}\label{actangder}
\Delta_{\Omega}\Ycksi(\theta,\varphi)=-\bigg(\xi^{2}+\frac{1}{4}\bigg)\Ycksi(\theta,\varphi)~,
\end{equation}
where $\Delta_{\Omega}$ is the differential operator
\begin{equation}\label{angder}
\Delta_{\Omega}=\frac{1}{\sinh\theta}\partial_{\theta}\big(\sinh\theta\partial_{\theta}\big)+
\frac{1}{\sinh^{2}\theta}\partial_{\varphi}^{2}~.
\end{equation}
Substituting the partial wave decomposition of $\phi$ in the free
part of the action we get
\begin{equation}\label{actiondecomp}
\begin{split}
S_{free}=-\frac{1}{2}\int dtdrd\theta d\varphi
r^{2}\sinh\theta\Biggr\{
\bigg(\sum_{m'=-\infty}^{+\infty}\frac{R_{\xi'm'}}{r}&\Ycksip\bigg)^{*}
\bigg[-\frac{1}{f}\partial_{t}^{2}+\frac{1}{r^{2}}\partial_{r}\big(r^{2}f\partial_{r}\big)\\
&+\frac{1}{r^{2}}\Delta_{\Omega}\bigg]\bigg(\sum_{m=-\infty}^{+\infty}
\frac{R_{\xi m}}{r} \Ycksi\bigg)\Biggl\}~,
\end{split}
\end{equation}
and with the help of the property (\ref{actangder})
\begin{equation}
\begin{split}
S_{free}=-\frac{1}{2}\sum_{m,m'}\int & dtdrd\theta d\varphi
\sinh\theta
\Biggr[R_{\xi'm'}^{*}\bigg(-\frac{1}{f}\bigg)\partial_{t}^{2}R_{\xi
m}+\frac{R_{\xi'm'}^{*}}{r}\partial_{r}\bigg(r^2f\partial_{r}\bigg(\frac{R_{\xi m}}{r}\bigg)\bigg)\\
&-\frac{R_{\xi'm'}^{*}}{r}\frac{R_{\xi
m}}{r}\bigg(\xi^{2}+\frac{1}{4}\bigg)\Biggr]\Ycksi(\theta,\varphi)\Big(\Ycksip(\theta,\varphi)\Big)^{*}~.
\end{split}
\end{equation}
Performing the integrations on $\theta$ and $\varphi$, using the
normalization condition (\ref{normcond}), we have
\begin{equation}
\begin{split}\label{rtaction}
S_{free}=-\frac{1}{2}\sum_{m=-\infty}^{\infty}\int dtdr\Biggr[R_{\xi
m}^{*}&\bigg(-\frac{1}{f}\bigg)\partial_{t}^{2}R_{\xi m}
+\frac{R_{\xi m}^{*}}{r}\partial_{r}\bigg(r^2f\partial_{r}\bigg(\frac{R_{\xi m}}{r}\bigg)\bigg)\\
&-\frac{R_{\xi m}^{*}}{r} \frac{R_{\xi
m}}{r}\bigg(\xi^{2}+\frac{1}{4}\bigg)\Biggr]~.
\end{split}
\end{equation}
Next, we will make a transformation to the tortoise coordinates
$(t,r_{*})$, defined by (\ref{tortoise}) and consider only the
region near the event horizon. In the case of
$-\frac{1}{3\sqrt{3}}<\mu<0$, this is the outer horizon $r_{+}$
and in the case of $\mu\geq 0$, this is the horizon $r_{h}$. In
order to include in our analysis both of these cases we denote
both $r_{+}$ and $r_{h}$ as $r_{H}$.

But first, we will determine the behaviour of the radial coordinate
$r$ and of the metric function $f(r)$ in tortoise coordinates in the
region near the horizon $r_{H}$. The Taylor's expansion of the
metric function $f(r)$ around the event horizon is
\begin{equation}
f(r)=2\kappa(r-r_H)+\sum_{n=2}^{\infty}\frac{f^{(n)}(r_H)}{n!}(r-r_H)^{n}~,
\end{equation}
where $\kappa\equiv\frac{1}{2}(\partial_r f)|_{r_{H}}$ is the
surface gravity. In the region near the event horizon we can keep
only the first two terms of the Taylor's expansion
\begin{equation}\label{Taylor2}
f(r)\approx2\kappa(r-r_H)~.
\end{equation}
Transforming to the tortoise coordinates $(t,r_\ast)$ and
integrating both sides of (\ref{tortoise}) using the approximation
(\ref{Taylor2}) we get
\begin{equation}
r_\ast\approx\int\frac{1}{2\kappa (r-r_H)}dr+C~,
\end{equation}
and so
\begin{equation}\label{rintort}
r(r_\ast)\approx A e^{2\kappa r_\ast}+r_H~,
\end{equation}
where $C$ is an arbitrary integration constant and $A\equiv
e^{-2\kappa C}$. Finally, the equations (\ref{Taylor2}) and
(\ref{rintort}) give
\begin{equation}\label{fintort}
f(r(r_\ast))\approx 2\kappa Ae^{2\kappa r_\ast}~.
\end{equation}
The last two equations describe the behaviour of $r$ and $f(r)$ in
tortoise coordinates in the region near the event  horizon. Note
that the limit $r\rightarrow r_H$ is equivalent to the limit
$r_\ast\rightarrow-\infty$ in tortoise coordinates, which means
that the event horizon in tortoise coordinates is located at
$(-\infty)$. In addition, note that near the event horizon
$f(r(r_\ast))$ vanishes exponentially fast, hence $f(r(r_\ast))$
is a suppression factor near the event horizon.

Now, we can return to the equation (\ref{rtaction}), transform to
tortoise coordinates and consider only the region near the horizon.
After using the fact that $f(r(r_{*}))$ is a suppression factor near
the horizon and keeping only dominant terms, the free part of the
action becomes in tortoise coordinates and in the region near the
horizon
\begin{equation}
(S_{free})_{*}=\sum_{m=-\infty}^{\infty}-\frac{1}{2}\int
dtdr_{*}R_{\xi m}^{*}
\bigg[-\partial_{t}^{2}+f\partial_{r}\big(f\partial_{r}\big)\bigg]
R_{\xi m}~,
\end{equation}
where the upper star denotes complex conjugation, the lower star
denotes the tortoise coordinates and $f, R_{\xi m}, R_{\xi m}^{*}$
are implicit functions of $r_{*}$. Transforming back to the original
coordinates we find
\begin{equation}\label{Sfreered}
S_{free}=\sum_{m=-\infty}^{\infty}-\frac{1}{2}\int dtdr R_{\xi
m}^{*} \bigg[-\frac{1}{f}\partial_{t}^{2}
+\partial_{r}\big(f\partial_{r}\big)\bigg]R_{\xi m}~.
\end{equation}
Concerning the part $S_{int}$ of the action, which includes a mass
term, potential terms and interaction terms, after performing a
partial wave decomposition in terms of the functions $\Ycksi$ and
upon transforming to the tortoise coordinates, one finds that all of
its terms contain the factor $f(r(r_{*}))$ and vanish exponentially
fast near the horizon. Thus, the total action $S$ is obtained
\begin{equation}
S=\sum_{m=-\infty}^{\infty}-\frac{1}{2}\int dtdr R_{\xi m}^{*}
\bigg[-\frac{1}{f}\partial_{t}^{2}
+\partial_{r}\big(f\partial_{r}\big)\bigg]R_{\xi m}~.
\end{equation}
According to this action, physics in the region near the horizon can
be effectively described by an infinite collection of
$(1+1)$-dimensional free massless complex scalar fields, each
propagating in a $(1+1)$-dimensional spacetime, which is given by
the $(t,r)$ part of the (3+1)-dimensional metric of the topological
black hole of genus $\gen=2$, that is
\begin{equation}\label{lineelred}
ds^{2}=-f(r)dt^{2}+\frac{1}{f(r)}dr^{2}~.
\end{equation}


\section{Hawking Radiation from Topological Black Holes}

In the reduced (1+1)-dimensional background (\ref{lineelred})
outgoing modes of the (1+1)-dimensional fields near the horizon
behave as right moving modes, while ingoing modes as left moving
modes. If we neglect the ingoing modes in the region near the
horizon, because they can not classically affect physics outside the
horizon, then the effective two-dimensional theory becomes chiral.
As it is known~\cite{AlvarezGaume:1983ig, Bardeen:1984pm,
Bertlmann:2000da, Bertlmann:2000book, Fujikawa:2004book} a
two-dimensional chiral theory exhibits a gravitational anomaly. The
consistent gravitational anomaly for right-handed fields
reads~\cite{AlvarezGaume:1983ig, Bertlmann:2000da}
\begin{equation}
\nabla_{\mu}T^{\mu}_{\nu}=\frac{1}{96\pi\sqrt{-g_{(2)}}}\epsilon^{\beta\delta}
\partial_{\delta}\partial_{\alpha}\Gamma^{\alpha}_{\nu\beta}~,
\end{equation}
and the covariant gravitational anomaly takes the form
\begin{equation}\label{covanom}
\nabla_{\mu}\Tcov^{\mu\nu}=\frac{\epsilon^{\nu\mu}}{96\pi\sqrt{-g_{(2)}}}\partial_{\mu}R~,
\end{equation}
where $T^{\mu}_{\nu}$ and $\Tcov^{\mu}_{\nu}$ are the consistent and
covariant energy-momentum tensor respectively,
$\epsilon^{01}=-\epsilon^{10}=1$, $R$ and $g_{(2)}$ are the Ricci
scalar and the metric determinant of the reduced metric
(\ref{lineelred}) respectively. The consistent gravitational anomaly
satisfies the Wess-Zumino consistency condition, but the consistent
energy-momentum tensor $T^{\mu}_{\nu}$ does not transform
covariantly under general coordinate transformations. The covariant
energy momentum tensor $\Tcov^{\mu}_{\nu}$, on the contrary,
transforms covariantly under general coordinate transformations, but
the covariant gravitational anomaly does not satisfy the Wess-Zumino
consistency condition. Consistent and covariant expressions are
related by local counterterms~\cite{Bardeen:1984pm,
Bertlmann:2000book, Fujikawa:2004book}. In~\cite{Iso:2006wa,
Iso:2006ut} the consistent expressions for the anomalies were taken,
whereas the imposed boundary conditions involved the covariant form.
A reformulation of this approach was given
in~\cite{Banerjee:2007qs}, where only covariant expressions were
used, rectifying this conceptual issue. Furthermore, a more
technically simplified way to obtain the Hawking flux was suggested
in~\cite{Banerjee:2008az,Banerjee:2007uc,Banerjee:2008wq}, where the
calculation involved only the expressions for the anomalous
covariant Ward identities and the covariant boundary conditions. We
will follow this approach to derive the Hawking flux.

We consider the expression for the two-dimensional covariant
gravitational Ward identity, that is the covariant anomaly
(\ref{covanom}), and taking its $\nu=t$ component we get
\begin{equation}\label{Wardt}
\partial_{r}\Tcov^{r}_{t}=\frac{1}{96\pi}f\partial_{r}f''~,
\end{equation}
where we have used the facts that the background is static and that
the Ricci scalar is $R=-f''(r)$, while a prime denotes
differentiation with respect to $r$. The equation (\ref{Wardt}) can
be written as
\begin{equation}\label{WardNrt}
\partial_{r}\Tcov^{r}_{t}=\partial_{r}\tilde{N}^{r}_{t}~,
\end{equation}
or
\begin{equation}\label{WardNrt2}
\partial_{r}\left(\Tcov^{r}_{t}-\tilde{N}^{r}_{t}\right)=0~,
\end{equation}
where
\begin{equation}\label{Nrt}
\tilde{N}^{r}_{t}=\frac{1}{96\pi}\left(
ff''-\frac{f'^{2}}{2}\right)~.
\end{equation}
Solving the equation (\ref{WardNrt}) we find
\begin{equation}\label{Trta}
\Tcov^{r}_{t}(r)=a_{H}+\tilde{N}^{r}_{t}(r)-\tilde{N}^{r}_{t}(r_{H})~.
\end{equation}
Here $a_{H}$ is an integration constant. Imposing the covariant
boundary condition~\cite{Iso:2006ut,Banerjee:2008wq}
\begin{equation}\label{bc}
\Tcov^{r}_{t}(r_{H})=0~,
\end{equation}
namely the vanishing of the covariant energy-momentum tensor at the
event horizon, yields  $a_{H}=0$. Hence, the anomalous covariant
energy-momentum tensor (\ref{Trta}) is
\begin{equation}\label{Trt}
\Tcov^{r}_{t}(r)=\tilde{N}^{r}_{t}(r)-\tilde{N}^{r}_{t}(r_{H})~.
\end{equation}
At what follows we restore in our formulae the value of the AdS
radius $l$, which had been set $l=1$. So, the metric function is
\begin{equation}
f(r)=\frac{r^{2}}{l^{2}}-1-\frac{2\mu}{r}~.
\end{equation}
We remind that the Hawking flux is measured at infinity, where there
is no gravitational anomaly and in~\cite{Robinson:2005pd,
Iso:2006wa, Iso:2006ut} it was given by the anomaly free (or
conserved) energy-momentum tensor. This required
in~\cite{Robinson:2005pd, Iso:2006wa, Iso:2006ut} to split the space
into two distinct regions, one near the horizon and the other away
from it, and use both the anomalous Ward identity in the vicinity of
the horizon and the normal Ward identity in the exterior region.
This is redundant if one observes that for the metric
(\ref{lineelred}) and the specific metric function $f(r)$ of the
(3+1)-dimensional topological black hole of genus $\gen=2$, the
gravitational anomaly vanishes at asymptotic infinity
$r\rightarrow\infty$. Indeed, we see that in this limit
\begin{equation}
\frac{\epsilon^{\nu\mu}}{96\pi\sqrt{-g}}\partial_{\mu}R=
-\frac{\epsilon^{\nu
1}}{96\pi}f'''=-\frac{\epsilon^{\nu1}}{96\pi}\frac{12\mu}{r^{4}}\longrightarrow
0~,
\end{equation}
and
\begin{equation}
\partial_{r}\tilde{N}^{r}_{t}=
\frac{1}{96\pi}\partial_{r}\left(ff''-\frac{f'^{2}}{2}\right)=
\frac{1}{96\pi}f\partial_{r}f''=
\frac{1}{96\pi}\left(\frac{12\mu}{l^{2}r^{2}}-\frac{12\mu}{r^{4}}-\frac{24\mu^{2}}{r^{5}}\right)\longrightarrow
0~.
\end{equation}
It is also important to notice that although the gravitational
anomaly vanishes at infinity, the $\tilde{N}^{r}_{t}$ does not,
since
\begin{equation}
\tilde{N}^{r}_{t}(r\rightarrow\infty)=-\frac{l^{-2}}{48\pi}~,
\end{equation}
because the spacetime asymptotically is not flat but it is AdS. The
last three equations and observation of the equation
(\ref{WardNrt2}) imply that the anomaly free (or conserved)
energy-momentum tensor, which is the energy flux $\mathit{\Phi}$
measured at infinity, is given by
\begin{equation}\label{Trtcons}
\begin{split}
\mathit{\Phi}&=\Tcov^{r}_{t}(r\rightarrow\infty)-\tilde{N}^{r}_{t}(r\rightarrow\infty)\\
&=-\frac{l^{-2}}{48\pi}-\tilde{N}^{r}_{t}(r_{H})-\left(-\frac{l^{-2}}{48\pi}\right)\\
&=-\tilde{N}^{r}_{t}(r_{H})~.
\end{split}
\end{equation}
Thus\footnote{~In the case of an asymptotically flat spacetime
treated in~\cite{Banerjee:2008az,Banerjee:2007uc,Banerjee:2008wq} it
was $\tilde{N}^{r}_{t}(r\rightarrow\infty)=0$ and the gravitational
anomaly vanished at infinity, so that the energy flux was calculated
as $\mathit{\Phi}=\Tcov^{r}_{t}(r\rightarrow\infty)=
-\tilde{N}^{r}_{t}(r_{H})$. The difference in the case of the
(3+1)-dimensional TBH of genus $\gen=2$ is that although the
gravitational anomaly vanishes at infinity, the
$\tilde{N}^{r}_{t}(r\rightarrow\infty)$ is not zero, due to the fact
that the spacetime is asymptotically AdS, so one must take it into
consideration according to the equation (\ref{WardNrt2}), in order
to find the correct conserved energy-momentum tensor at infinity. Of
course, if we put $\tilde{N}^{r}_{t}(r\rightarrow\infty)=0$ in the
equations (\ref{WardNrt}), (\ref{WardNrt2}), (\ref{Trtcons}) and
(\ref{Trt}) we retrieve the result for the asymptotically flat case.
Note that finally for both the asymptotically AdS spacetime and the
asymptotically flat spacetime the energy flux is given by the
equation $\mathit{\Phi}=-\tilde{N}^{r}_{t}(r_{H}).$}, the energy
flux measured at infinity is
\begin{equation}
\mathit{\Phi}=-\tilde{N}^{r}_{t}(r_{H})=\frac{1}{192\pi}f'^{2}(r_{H})~,
\end{equation}
or in a different form
\begin{equation}\label{flux}
\mathit{\Phi}=\frac{\pi}{12}\left(\frac{f'(r_{H})}{4\pi}\right)^{2}~.
\end{equation}
A beam of massless blackbody radiation  moving in the positive $r$
direction at a temperature $T$ has a flux of the form
$\mathit{\Phi}=\frac{\pi}{12}T^{2}$. Therefore, we see that the flux
(\ref{flux}) has a form equivalent to blackbody radiation with a
temperature
\begin{equation}
T_{H}=\frac{f'(r_{H})}{4\pi}=\frac{\kappa}{2\pi}~.
\end{equation}
This temperature is exactly the Hawking temperature of a
(3+1)-dimensional topological black hole of genus $\gen=2$ as
determined in~\cite{Vanzo:1997gw}. Hence, $\mathit{\Phi}$ is the
Hawking flux.


\section{Hawking Radiation from a TBH Conformally Coupled to a Scalar Field }

Another interesting non-spherical background is a TBH conformally
coupled to a scalar field. Consider four-dimensional gravity with
negative cosmological constant $(\Lambda=-3l^{-2})$ and a scalar
field $\phi(x)$ described by the action
\begin{equation}\label{IEinframe}
I[g_{\mu\nu},\phi]=\int
d^{4}x\sqrt{-g}\biggl[\frac{R_{E}+6l^{-2}}{16\pi
G}-\frac{1}{2}g^{\mu\nu}\partial_{\mu}\phi\partial_{\nu}\phi-V(\phi)\biggr]~,
\end{equation}
where $R_{E}$ is the Ricci scalar in the Einstein frame, $l$ is the
AdS radius and $G$ is the Newton's constant. We take the following
self-interaction potential
\begin{equation}\label{poten}
V(\phi)=-\frac{3}{4\pi Gl^{2}}\sinh^{2}{\sqrt{\frac{4\pi
G}{3}}\phi}~.
\end{equation}
It was proved in~\cite{Martinez:2004nb} that there is a static
black hole solution (MTZ black hole) with topology
$\mathbb{R}^{2}\times\Sigma$, where $\Sigma$ is a two-dimensional
manifold of constant negative curvature, which is locally
isomorphic to the hyperbolic manifold $H^{2}$ and of the form
\begin{equation}
\Sigma=H^{2}/\Gamma \quad \textrm{,\quad  $\Gamma\subset O(2,1)$}~,
\end{equation}
where $\Gamma$ is a freely acting discrete subgroup (i.e. without
fixed points) of isometries. This black hole solution is given by
\begin{equation}\label{MTZEin}
ds^{2}=\frac{r(r+2G\mu)}{(r+G\mu)^{2}}\biggl[-\biggl(\frac{r^{2}}{l^{2}}-\biggl(1+\frac{G\mu}{r}\biggr)^2\biggr)dt^{2}+
\biggl(\frac{r^{2}}{l^{2}}-\biggl(1+\frac{G\mu}{r}\biggr)^{2}\biggr)^{-1}dr^{2}+r^{2}d\sigma^{2}\biggr]~,
\end{equation}
and the scalar field is
\begin{equation}\label{MTZEin.sc}
\phi(r)=\sqrt{\frac{3}{4\pi G}}\arctanh\frac{G\mu}{r+G\mu}~.
\end{equation}
Here $d\sigma^{2}$ is the line element of the two-dimensional
manifold $\Sigma$
\begin{equation}
d\sigma^{2}=d\theta^{2}+\sinh^{2}\theta d\varphi^{2}~,
\end{equation}
where $\theta\ge0$ and $0\le\varphi<2\pi$ are the coordinates of the
hyperbolic space $H^{2}$. The mass of this solution is given by
\begin{equation}
M=\frac{\sigma}{4\pi}\mu~,
\end{equation}
where $\sigma$ denotes the area of $\Sigma$ and $\mu>-l/4G$ is a
constant. Performing a conformal transformation with a scalar field
redefinition of the form
\begin{equation}
\begin{split}
\hat{g}_{\mu\nu}&=\biggl(1-\frac{4\pi G}{3}\Psi^{2}\biggr)^{-1}g_{\mu\nu}~,\\
\Psi&=\sqrt{\frac{3}{4\pi G}}\tanh{\sqrt{\frac{4\pi G}{3}}\phi}~,
\end{split}
\end{equation}
the action (\ref{IEinframe}) and (\ref{poten}) reads
\begin{equation}\label{Iconfframe}
I[\hat{g}_{\mu\nu},\Psi]=\int
d^{4}x\sqrt{-\hat{g}}\biggl[\frac{\hat{R}+6l^{-2}}{16\pi
G}-\frac{1}{2}\hat{g}^{\mu\nu}\partial_{\mu}\Psi\partial_{\nu}\Psi
-\frac{1}{12}\hat{R}\Psi^{2}-\frac{2\pi G}{3l^{2}}\Psi^{4}\biggr]~.
\end{equation}
In this frame the scalar field equation is conformally invariant,
since the matter action is invariant under arbitrary local
rescalings
$\hat{g}_{\mu\nu}\rightarrow\lambda^{2}(x)\hat{g}_{\mu\nu}$ and
$\Psi\rightarrow\lambda^{-1}\Psi$. The black hole solution
(\ref{MTZEin}) and (\ref{MTZEin.sc}) acquires a simple form once
expressed in the conformal frame
\begin{equation}\label{MTZconf}
d\hat{s}^{2}=-\biggl[\frac{r^{2}}{l^{2}}-\biggl(1+\frac{G\mu}{r}\biggr)^{2}\biggr]dt^{2}+
\biggl[\frac{r^{2}}{l^{2}}-\biggl(1+\frac{G\mu}{r}\biggr)^{2}\biggr]^{-1}dr^{2}+r^{2}d\sigma^{2}~,
\end{equation}
with
\begin{equation}\label{Psiconf}
\Psi(r)=\sqrt{\frac{3}{4\pi G}}\frac{G\mu}{r+G\mu}~.
\end{equation}
We define
\begin{equation}\label{fconf}
f(r)\equiv\frac{r^{2}}{l^{2}}-\biggl(1+\frac{G\mu}{r}\biggr)^{2}~,
\end{equation}
and the metric (\ref{MTZconf}) is written as
\begin{equation}
d\hat{s}^{2}=-f(r)dt^{2}+\frac{1}{f(r)}dr{^2}+r^{2}d\theta^{2}+r^{2}\sinh^{2}\theta
d\varphi^{2}~.
\end{equation}
We consider only the case in which the two-dimensional manifold
$\Sigma$ is a compact two-dimensional manifold of genus $\gen=2$,
with constant negative curvature, after the identifications that we
have mentioned in section 2. For non-negative mass $\mu\geq{0}$,
this solution possesses only one event horizon at
\begin{equation}
r_{+}=\frac{l}{2}\biggl(1+\sqrt{1+\frac{4G\mu}{l}}\biggr)~,
\end{equation}
and $\Psi$ is regular everywhere. For negative mass $-l/4<G\mu<0$,
the metric (\ref{MTZconf}) develops three horizons, two of which are
event horizons located at $r_{--}$ and at $r_{+}$
 \be
r_{--}=\frac{l}{2}\biggl(-1+\sqrt{1-\frac{4G\mu}{l}}\biggr)~,
 \ee
 \be
r_{-}=\frac{l}{2}\biggl(1-\sqrt{1+\frac{4G\mu}{l}}\biggr)~,
 \ee
 \be
r_{+}=\frac{l}{2}\biggl(1+\sqrt{1+\frac{4G\mu}{l}}\biggr)~,
 \ee
 \\
which satisfy $0<r_{--}<-G\mu<r_{-}<l/2<r_{+}$. The scalar field
$\Psi$ is singular at $r=-G\mu$. The Ricci scalar of the black hole
solution (\ref{MTZconf}) in the conformal frame is
\begin{equation}\label{RMTZ}
\hat{R}=-12 l^{-2}~.
\end{equation}

As before, we consider a complex scalar field $\hat{\phi}(x)$ in the
background of the MTZ black hole of genus $\gen=2$ with a scalar
hair $\Psi$, in the conformal frame. This field has an action of the
form
\begin{equation}
S=S_{free}+S_{int}~,
\end{equation}
where $S_{free}$ is the free part of the action
\begin{equation}\label{SfreeMTZ}
S_{free}=-\frac{1}{2}\int
d^{4}x\sqrt{-\hat{g}}\hat{\phi}^{*}\bigg[-\frac{1}{f}dt^{2}
+\frac{1}{r^{2}}\partial_{r}\big(r^{2}f\partial_{r}\big)+\frac{1}{r^{2}}\Delta_{\Omega}\bigg]\hat{\phi}~,
\end{equation}
and $S_{int}$ is the part of the action which includes a mass term,
potential terms and interaction terms, where we have ignored the
interaction of $\hat{\phi}$ with $\Psi$. We perform a partial wave
decomposition of $\hat{\phi}$ in terms of the functions $\Ycksi$
\begin{equation}
\hat{\phi}(t,r,\theta,\varphi)=\sum_{m=-\infty}^{+\infty}\frac{R_{\xi
m}(t,r)}{r}\Ycksi(\theta,\varphi)~.
\end{equation}
We substitute the partial wave decomposition in the free action and
transform to the tortoise coordinates $(t,r_{*})$ defined by
(\ref{tortoise}). Then, one finds in the region near the event
horizon $r_{+}$, which for $-l/4<G\mu<0$ is the outer event horizon
and for $\mu\geq{0}$ is the unique event horizon, that the effective
radial potentials for partial wave modes of the field contain the
suppression factor $f(r(r_{*}))$ and vanish exponentially fast. The
same applies to the mass terms and interaction terms of the part
$S_{int}$. Thus, physics in the region near the horizon can be
effectively described by an infinite collection of
$(1+1)$-dimensional free massless scalar fields, each propagating in
a $(1+1)$-dimensional spacetime, which is given by the $(t,r)$ part
of the (3+1)-dimensional metric of the MTZ black hole of genus
$\gen=2$ in the conformal frame, that is
\begin{equation}\label{lineelredMTZ}
d\hat{s}^{2}=-f(r)dt^{2}+\frac{1}{f(r)}dr^{2}~.
\end{equation}

In this two-dimensional background we identify outgoing modes near
the horizon as right moving modes, while ingoing modes as left
moving modes. Neglecting the classically irrelevant ingoing modes in
the region near the horizon, the effective two-dimensional theory
becomes chiral and a gravitational anomaly appears. The covariant
gravitational anomaly for right-handed fields reads
\begin{equation}\label{covanomMTZ}
\nabla_{\mu}\Tcov^{\mu\nu}=\frac{\epsilon^{\nu\mu}}{96\pi\sqrt{-\hat{g}_{(2)}}}\partial_{\mu}R~,
\end{equation}
where $\Tcov^{\mu}_{\nu}$ is the covariant energy-momentum tensor,
$\epsilon^{01}=-\epsilon^{10}=1$, $R=-f''(r)$ and $\hat{g}_{(2)}$
are the Ricci scalar and the metric determinant of the reduced
metric (\ref{lineelredMTZ}) respectively. Taking the $\nu=t$
component of the two-dimensional covariant anomaly
(\ref{covanomMTZ}), we have
\begin{equation}\label{WardtMTZ}
\partial_{r}\Tcov^{r}_{t}=\frac{1}{96\pi}f\partial_{r}f''~.
\end{equation}
This equation is written as
\begin{equation}\label{WardNrtMTZ}
\partial_{r}\Tcov^{r}_{t}=\partial_{r}\tilde{N}^{r}_{t}~,
\end{equation}
or
\begin{equation}\label{WardNrt2MTZ}
\partial_{r}\left(\Tcov^{r}_{t}-\tilde{N}^{r}_{t}\right)=0~,
\end{equation}
where
\begin{equation}\label{NrtMTZ}
\tilde{N}^{r}_{t}=\frac{1}{96\pi}
\left(ff''-\frac{f'^{2}}{2}\right)~.
\end{equation}
Solving the equation (\ref{WardNrtMTZ}) we find
\begin{equation}\label{TrtaMTZ}
\Tcov^{r}_{t}(r)=b_{+}+\tilde{N}^{r}_{t}(r)-\tilde{N}^{r}_{t}(r_{+})~.
\end{equation}
Here $b_{+}$ is an integration constant. Implementing the usual
covariant boundary condition
\begin{equation}\label{bcMTZ}
\Tcov^{r}_{t}(r_{+})=0~,
\end{equation}
yields $b_{+}=0$. Therefore, the anomalous covariant energy-momentum
tensor is
\begin{equation}\label{TrtMTZ}
\Tcov^{r}_{t}(r)=\tilde{N}^{r}_{t}(r)-\tilde{N}^{r}_{t}(r_{+})~.
\end{equation}
We notice that for the metric (\ref{lineelredMTZ}) with a metric
function $f(r)$ given by the equation (\ref{fconf}), the
gravitational anomaly vanishes at asymptotic infinity
$r\rightarrow\infty$, but the $\tilde{N}^{r}_{t}$ does not due to
the fact that asymptotically the spacetime is AdS. Indeed, in this
limit we have
\begin{equation}
\frac{\epsilon^{\nu\mu}}{96\pi\sqrt{-g}}\partial_{\mu}R=
-\frac{\epsilon^{\nu 1}}{96\pi}f'''=-\frac{\epsilon^{\nu 1}}{96\pi}
\left(\frac{12G\mu}{r^{4}}+\frac{24(G\mu)^{2}}{r^{5}}\right)
\longrightarrow 0~,
\end{equation}
and
\begin{equation}
\begin{split}
\partial_{r}\tilde{N}^{r}_{t}
=\frac{1}{96\pi}\biggr(\frac{12G\mu}{l^{2}r^{2}}+\frac{24(G\mu)^{2}}{l^{2}r^{3}}
&-\frac{12G\mu}{r^{4}}-\frac{48(G\mu)^{2}}{r^{5}}\\
&-\frac{60(G\mu)^{3}}{r^{6}}-\frac{24(G\mu)^{4}}{r^{7}}\biggl)\longrightarrow
0~,
\end{split}
\end{equation}
but
\begin{equation}
\tilde{N}^{r}_{t}(r\rightarrow\infty)=-\frac{l^{-2}}{48\pi}~.
\end{equation}
Using the same arguments as in section 4 we see that the anomaly
free energy-momentum tensor, and thus the energy flux
$\mathit{\Phi}$ measured at infinity, is
\begin{equation}\label{fluxMTZ}
\begin{split}
\mathit{\Phi}&=
\Tcov^{r}_{t}(r\rightarrow\infty)-\tilde{N}^{r}_{t}(r\rightarrow\infty)\\
&=-\frac{l^{-2}}{48\pi}-\tilde{N}^{r}_{t}(r_{+})-\left(-\frac{l^{-2}}{48\pi}\right)\\
&=-\tilde{N}^{r}_{t}(r_{+})\\
&=\frac{1}{192\pi}f'^{2}(r_{+})~,
\end{split}
\end{equation}
or in a different form
\begin{equation}
\mathit{\Phi}=\frac{\pi}{12}\left(\frac{f'(r_{+})}{4\pi}\right)^{2}~.
\end{equation}
This flux is equivalent to a flux of blackbody radiation with a
temperature
\begin{equation}
T_{H}=\frac{f'(r_{+})}{4\pi}=\frac{\kappa}{2\pi}~,
\end{equation}
where $\kappa\equiv\frac{1}{2}(\partial_r f)|_{r_{+}}$ is the
surface gravity. The temperature $T_{H}$ is identical to the Hawking
temperature of the MTZ black hole as determined
in~\cite{Nadalini:2007qi, Martinez:2005di}. Hence, $\mathit{\Phi}$
is identified with the Hawking flux of the MTZ black hole, which is
a (3+1)-dimensional topological black hole conformally coupled to a
scalar field.


\section{Robinson-Wilczek Method with a Scalar Field Non-minimally Coupled to the Black Hole Background
}

In the previous section we showed  that the scalar field $\Psi$,
which is coupled to the black hole does not explicitly contribute to
the Hawking radiation. The reason is that the scalar field does not
introduce any new conserved charge and its only effect is to alter
the form of the background black hole solution to a maximal
Reissner-Nordstr\"om-AdS black hole. In the Robinson-Wilczek method
this is expected since the scalar field is time-independent and
therefore it can not generate a flux. For this reason, if we had
tried to perform the usual reduction procedure only with the scalar
field $\Psi$, assuming that it gives Hawking radiation, we would
have found that its action in the vicinity of the event horizon
vanishes due to the suppression factor $f(r(r_{*}))$ (see Appendix
B). However, it is interesting to investigate what happens if the
scalar field, which parametrizes the matter, backreacts on the
geometry. In this direction we will discuss the consequences that
possibly occur to the standard Robinson-Wilczek method if this
scalar field is non-minimally coupled to gravity.

We consider for simplicity a static spherically symmetric
four-dimensional spacetime
\begin{equation}\label{genBH}
ds^{2}=-f(r)dt^{2}+\frac{1}{f(r)}dr{^2}+r^{2}d\theta^{2}+r^{2}\sin^{2}\theta
d\varphi^{2}~,
\end{equation}
where $f(r)$   is a function which admits at least one event
horizon. The event horizon is located at $r=r_H$ where $f(r_H)=0$
and the surface gravity is $\kappa\equiv\frac{1}{2}(\partial_r
f)|_{r_H}$. We also consider an interacting scalar field $\phi(x)$,
which is non-minimally coupled to the black hole background
(\ref{genBH}). The action of this scalar field is
\begin{equation}\label{genaction}
S\left[\phi\right]=\frac{1}{2}\int
d^{4}x\sqrt{-g}\biggl[g^{\mu\nu}\partial_{\mu}\phi\partial_{\nu}\phi-\sum_{n=2}^\infty
\lambda_n \phi^n-\alpha R \phi^{2} \biggr]~,
\end{equation}
where $\lambda_n$ are a set of arbitrary coupling constants (for
example $\lambda_{2}\equiv m^{2}$ gives the mass), $R$ is the Ricci
scalar and $\alpha$ is a coupling constant to gravity. In particular
\begin{equation}
\alpha=\begin{cases}
 0& \text{for minimally coupled $\phi(x)~,$} \\
 \frac{D-2}{4(D-1)}& \text{for conformally coupled $\phi(x)~,$}
 \end{cases}
\end{equation}
for a $D$-dimensional spacetime. For the spacetime (\ref{genBH}) it
is $D=4$ and $\alpha=1/6$, if $\phi(x)$ is conformally coupled to
the black hole background. We can write the action (\ref{genaction})
as the sum of three different terms, each having a different
physical meaning
\begin{equation}
S=S_{free}+S_{int}+S_{c}~,
\end{equation}
where the first term is
\begin{equation}
S_{free}=\frac{1}{2}\int
d^{4}x\sqrt{-g}g^{\mu\nu}\partial_{\mu}\phi\partial_{\nu}\phi=-\frac{1}{2}\int
d^{4}x\sqrt{-g}\phi\nabla^{2}\phi~,
\end{equation}
and $\nabla^{2}$ is the Laplace-Beltrami operator. The part
$S_{free}$ is the free part of the action. The second term is
\begin{equation}
S_{int}=-\frac{1}{2}\int d^{4}x\sqrt{-g}\sum_{n=2}^\infty \lambda_n
\phi^n~,
\end{equation}
and describes the interactions of the scalar field. The third term
is
\begin{equation}\label{Sgrav}
S_{c}=-\frac{1}{2}\int d^{4}x\sqrt{-g}\alpha R\phi^{2}~,
\end{equation}
and it is the part of the action $S$ which describes the non-minimal
coupling of the scalar field to the black hole background
(\ref{genBH}). We will mainly focus our attention on $S_{c}$. The
Ricci scalar is
\begin{equation}\label{Rnonmc}
R=-f''(r)-\frac{4f'(r)}{r}-\frac{2f(r)}{r^{2}}+\frac{2}{r^{2}}~,
\end{equation}
where a prime denotes differentiation with respect to $r$. The
partial wave decomposition of the scalar field in terms of the
spherical harmonics is
\begin{equation}\label{expand}
\phi(x)=\sum_{l,m}\frac{u_{lm}(t,r)}{r}Y_{l}^{m}(\theta,\varphi)~,
\end{equation}
and substituting to the action (\ref{Sgrav}), we find, after
performing the integrations on $\theta,\varphi$ with the help of the
normalization and orthogonality conditions of the spherical
harmonics, that
\begin{equation}
S_{c}=-\frac{1}{2}\int dtdr\alpha
R\sum_{l_{1},l_{2}}\sum_{m_{1},m{2}}u_{l_{1}m{_1}}u_{l_{2}m{_2}}\delta_{l_{1}l{_2}}\delta_{m_{1}m{_2}}~,
\end{equation}
Using the expression (\ref{Rnonmc}) for $R$ we get
\begin{equation}
S_{c}=\frac{1}{2}\int
dtdr\alpha\left[f''(r)+\frac{4f'(r)}{r}+\frac{2f(r)}{r^{2}}-\frac{2}{r^{2}}\right]
\sum_{l_{1},l_{2}}\sum_{m_{1},m{2}}u_{l_{1}m{_1}}u_{l_{2}m{_2}}{^{(2)}C_{\{l_{1},l_{2}\}}^{\{m_{1},m{2}\}}}~,
\end{equation}
where
$^{(2)}C_{\{l_{1},l_{2}\}}^{\{m_{1},m{2}\}}\equiv\delta_{l_{1}l{_2}}\delta_{m_{1}m{_2}}$.
A transformation to tortoise coordinates $(t,r_*)$, defined by the
equation (\ref{tortoise}), transforms $S_{c}$ to
\begin{equation}
\begin{split}\label{Sgrintort}
S_{c\ast}=\frac{1}{2}\int dtdr_\ast &\Biggl\{ f(r(r_\ast))\alpha
\biggl[f''(r(r_\ast))+\frac{4f'(r(r_\ast))}{r(r_\ast)}+\frac{2f(r(r_\ast))}{r^{2}(r_\ast)}-\frac{2}{r^{2}(r_\ast)}\biggr]\\
&\times\sum_{l_{1},l_{2}}\sum_{m_{1},m{2}}u_{l_{1}m{_1}}u_{l_{2}m{_2}}{^{(2)}
C_{\{l_{1},l_{2}\}}^{\{m_{1},m{2}\}}}\Biggr\}~,
\end{split}
\end{equation}
where now $r,f(r),u_{l_{1}m{_1}},u_{l_{2}m{_2}}$ are thought as
implicit functions of $r_\ast$ and the prime still denotes
differentiation with respect to $r$. In the region near the event
horizon we have proved that
\begin{equation}\label{rintortc}
r(r_\ast)\approx A e^{2\kappa r_\ast}+r_H~,
\end{equation}
and
\begin{equation}\label{fintortc}
f(r(r_\ast))\approx 2\kappa Ae^{2\kappa r_\ast}~.
\end{equation}
Hence, the limit $r\rightarrow r_H$ is equivalent to
$r_\ast\rightarrow-\infty$ in tortoise coordinates and
$f(r(r_\ast))$ is a suppression factor near the horizon. Now, we
examine how each term of $S_{c\ast}$ behaves in the vicinity of the
horizon, using the equations (\ref{rintortc}) and (\ref{fintortc}),
in order to find which terms are dominant. We easily see that in
this region
\begin{equation}\label{1termSgr}
f(r(r_\ast))\frac{2f(r(r_\ast))}{r^{2}(r_\ast)}\,\longrightarrow{0}~,
\end{equation}
\begin{equation}\label{2termSgr}
-f(r(r_\ast))\frac{2}{r^{2}(r_\ast)}\:\longrightarrow{0}~.
\end{equation}
The other two terms need special attention, so we write for the
region near the event horizon
\begin{equation}\label{derterm1}
\begin{split}
f'(r(r_\ast))&=\frac{\partial f(r(r_\ast))}{\partial
r}=\frac{\partial f(r(r_\ast))}{\partial r_\ast}\frac{\partial
r_\ast}{\partial r}\\
\\
&=\frac{\partial f(r(r_\ast))}{\partial
r_\ast}\frac{1}{f(r(r_\ast))}=\frac{\partial}{\partial
r_\ast}\big[\ln{ f(r(r_\ast))}\big]\approx 2\kappa~.
\end{split}
\end{equation}
Thus, we find for $r_{\ast}\rightarrow-\infty$
\begin{equation}\label{3termSgr}
f(r(r_\ast))\frac{4f'(r(r_\ast))}{r((r_\ast))}\,\longrightarrow{0}~.
\end{equation}
Similarly, we write
\begin{equation}\label{derterm2}
\begin{split}
f(r(r_\ast))\frac{\partial ^{2}f(r(r_\ast))}{\partial
r^{2}}&=f(r(r_\ast))\frac{\partial}{\partial
r}\biggl[\frac{\partial}{\partial r_\ast}\ln{
f(r(r_\ast))}\biggr]\\
&=f(r(r_\ast))\frac{\partial}{\partial
r_\ast}\biggl[\frac{\partial}{\partial r_\ast}\ln{
f(r(r_\ast))}\biggr]\frac{\partial r_\ast}{\partial
r}\\
&=f(r(r_\ast))\frac{\partial}{\partial
r_\ast}\biggl[\frac{\partial}{\partial r_\ast}\ln{
f(r(r_\ast))}\biggr]\frac{1}{f(r(r_\ast))}\\
&=\frac{\partial}{\partial r_\ast}\biggl[\frac{\partial}{\partial
r_\ast}\ln{ f(r(r_\ast))}\biggr]~.
\end{split}
\end{equation}
Hence, we get
\begin{equation}\label{4termSgr}
f(r(r_\ast))f''(r(r_\ast))\,\longrightarrow{0}~.
\end{equation}
From the equations (\ref{1termSgr}), (\ref{2termSgr}),
(\ref{3termSgr}) and (\ref{4termSgr}) the action (\ref{Sgrintort})
in the region near the event horizon becomes $S_{c\ast}={0}$ and
therefore
\begin{equation}\label{SgrH}
S_{c}={0}~.
\end{equation}
Regarding the part $S_{int}$ of the total action, which describes
the interactions of the scalar field $\phi(x)$, after performing a
partial wave decomposition of $\phi(x)$ in terms of the spherical
harmonics and upon transforming to the tortoise coordinates, one
finds~\cite{Robinson:2005pd, Iso:2006wa, Iso:2006ut} that it
vanishes exponentially fast near the event horizon
\begin{equation}\label{SintH}
S_{int}={0}~,
\end{equation}
due to the presence of the suppression factor $f(r(r_{*}))$.
Concerning the free part $S_{free}$ of the total action, after
performing a partial wave decomposition of $\phi(x)$ of the form of
(\ref{expand}), transforming to the tortoise coordinates and keeping
only dominant terms~\cite{Robinson:2005pd, Iso:2006wa, Iso:2006ut}
we find in the region near the event horizon
\begin{equation}\label{SfreeH}
S_{free}=-\frac{1}{2}\sum_{l,m}\int
dtdru_{lm}\left[-\frac{1}{f}\partial_{t}^{2}+\partial_{r}(f\partial_{r})\right]u_{lm}~.
\end{equation}
Adding the equations (\ref{SgrH}), (\ref{SintH}) and (\ref{SfreeH})
we get the total action for the region near the event horizon
\begin{equation}
S=\sum_{l,m}-\frac{1}{2}\int
dtdru_{lm}\left[-\frac{1}{f}\partial_{t}^{2}+\partial_{r}(f\partial_{r})\right]u_{lm}~.
\end{equation}
Thus, physics near the horizon can be described using an infinite
set of (1+1)-dimensional massless scalar fields, each propagating in
a (1+1)-dimensional background with a metric
\begin{equation}
ds^{2}=-f(r)dt^{2}+\frac{1}{f(r)}dr{^2}~.
\end{equation}

In conclusion, the non-minimal coupling of the scalar field to the
gravitational background does not introduce any special modification
to the reduction procedure, since the part of the action $S_{c}$,
which describes this non-minimal coupling, vanishes in the region
near the event horizon. Then, the standard Robinson-Wilczek method
proceeds in exactly the same way as in the case of a minimally
coupled scalar field. Of course, the preceding analysis can be
generalized for $D$-dimensional spacetimes $(D>4)$, which have a
metric of the type
\begin{equation}
ds^{2}=-f(r)dt^{2}+\frac{1}{f(r)}dr{^2}+r^{2}d\Omega^{2}_{D-2}~,
\end{equation}
with the difference that we must expand the scalar field in terms of
the $(D-2)$-dimensional spherical harmonics and integrate over a
$(D-2)$-dimensional sphere.

We should clarify one point here. The wave equation of a scalar
field minimally coupled or non-minimally coupled to gravity in
higher than two dimensions will develop a potential which away from
the horizon will modify the Hawking radiation. Therefore, the actual
Hawking radiation observed at infinity is calculated through the
grey-body factors. However, in the Robinson-Wilczek method  the
thermal Hawking flux results from the infinite (1+1)-dimensional
fields which act as the thermal source of this flux.


\section{Summary}

We studied the method of calculating the Hawking radiation via
gravitational anomalies in gravitational backgrounds of constant
negative curvature. At first we discussed the mode analysis of the
scalar wave equation in the background of a topological black hole.
In the case of (3+1)-dimensional topological black holes of genus
$\gen=2$, we performed the dimensional reduction procedure to two
dimensions and we showed that near the horizon the matter scalar
field is reduced to an infinite collection of (1+1)-dimensional free
massless scalar fields. To calculate the Hawking radiation from the
topological black holes of genus $\gen=2$ we followed the covariant
anomalies approach proposed
in~\cite{Banerjee:2008az,Banerjee:2007uc,Banerjee:2008wq}, which we
modified in order to include asymptotically non-flat spacetimes,
because it is conceptually simpler and technical problems connected
with a complicated horizon structure of the topological black holes
of genus $\gen=2$ can be avoided.

We also applied this method to a (3+1)-dimensional topological black
hole of genus $\gen=2$ conformally coupled to a scalar field and we
retrieved the correct Hawking flux and temperature. These solutions
are interesting because they are examples of a scalar field
backreacting on the geometry. Because the scalar field is static it
can not give an extra contribution to the Hawking flux. However,
there exist solutions of BTZ-type black holes coupled to
time-dependent scalar fields~\cite{Bak:2007jm}. These solutions are
not analytical so it is not clear how the Robinson-Wilczek method
can be applied to these backgrounds.

It is interesting to investigate if the Robinson-Wilczek method
can be applied to general backgrounds where the scalar field
responsible for the Hawking flux backreacts on the geometry. In
this direction, we addressed the problem of using in the method of
gravitational anomalies a scalar field non-minimally coupled to
the gravitational background, instead of a minimally coupled
scalar field as it is customary. We proved explicitly that the
non-minimal coupling does not affect the dimensional reduction
procedure and the method in general, since the part of the action
which describes the non-minimal coupling vanishes in the region
near the event horizon. Another interesting problem to address is
to examine the applicability of the gravitational anomaly method
in fully dynamical backgrounds, but one has first to tackle more
fundamental problems like how one can apply the technique of
dimensional reduction to time-dependent backgrounds and how one
can define uniquely the surface gravity for time-dependent
horizons (for a recent discussion on dynamical black holes
see~\cite{Vanzo:2008uq}).

\section*{Acknowledgments}

This work was supported by the NTUA research program PEVE07.

\vspace{1.0cm}

\appendix

\section{Appendix: The Functions $\Ycksi$}

The functions $\mathcal{P}_{l}^{mn}(z)$~\cite{Balazs:1986uj,
Argyres:1988ga} form the canonical basis for the irreducible
representations of the group SL$(2,C)$ and can be viewed as
playing the same role for the group SU$(1,1)$. They are defined in
the complex $z$ plane with a cut located on the real axis between
$-1,+1$. A convenient representation, which can serve as a
definition for the functions $\mathcal{P}_{l}^{mn}(z)$, is
\begin{equation}
\Pcmn=\frac{1}{2\pi
i}\int_{C}dz\left(\cosh\frac{\theta}{2}-z\sinh\frac{\theta}{2}\right)^{l+n}
\left(\sinh\frac{\theta}{2}+z\cosh\frac{\theta}{2}\right)^{l-n}z^{m-l-1}~,
\end{equation}
where $C$ is the unit circle prescribed positively, $m$ and $n$
are integers and $l$ can be complex (typically of the form
$l=-\frac{1}{2}\pm i\xi$, $\xi>0$). The generating function of the
$\mathcal{P}_{l}^{mn}$ is
\begin{equation}
\begin{split}
\sum_{m=-\infty}^{\infty}\Pcmn
&e^{-im\varphi}\\
=e^{-in\varphi}&\left(\cosh\frac{\theta}{2}+e^{i\varphi}\sinh\frac{\theta}{2}\right)^{l+n}
\left(\cosh\frac{\theta}{2}+e^{-i\varphi}\sinh\frac{\theta}{2}\right)^{l-n}~.
\end{split}
\end{equation}
In the case of $n=0$, we have
\begin{equation}
\sum_{m=-\infty}^{\infty}\mathcal{P}^{m0}_{-\frac{1}{2}\pm
i\xi}(\cosh\theta)
e^{-im\varphi}=\left(\cosh\theta+\sinh\theta\cos\varphi\right)^{-\frac{1}{2}\pm
i\xi}~.
\end{equation}
They have the following properties
\begin{eqnarray}
\Pcmn&=&\mathcal{P}_{l}^{-m,-n}(\cosh\theta)~,\\
\Pcmn&=&(-1)^{m-n}\mathcal{P}_{-l-1}^{nm}(\cosh\theta)~,\\
\big[\Pcmn\big]^{*}&=&\mathcal{P}_{l^{*}}^{mn}(\cosh\theta)~.
\end{eqnarray}
The functions $\Pc$ are related to the associated Legendre functions
$P_{l}^{m}$ through
\begin{equation}
\Pc(\cosh\theta)=\frac{\Gamma(l+1)}{\Gamma(l+m+1)}P_{l}^{m}(\cosh\theta)~,
\end{equation}
and for $l=-\frac{1}{2}+i\xi$, this is
\begin{equation}
\Pcksi(\cosh\theta)=\frac{\Gamma(i\xi+\frac{1}{2})}{\Gamma(i\xi+m+\frac{1}{2})}
P_{-\frac{1}{2}+i\xi}^{m}(\cosh\theta)~.
\end{equation}
The functions $\Pcksi$ form a complete set of functions on the
pseudosphere and satisfy the orthogonality relation
\begin{equation}\label{orthoPcksi}
\int_{0}^{\infty}d\theta\sinh\theta\Pcksi(\cosh\theta)
\Big(\mathcal{P}^{m0}_{-\frac{1}{2}+i\xi'}(\cosh\theta)\Big)^{*}=\frac{1}{4\pi^{2}}\xi\tanh(\pi\xi)\delta(\xi-\xi')~,
\end{equation}
where $\xi,\xi'\geq0$. When $\xi,\xi'$ take discrete real values the
delta function $\delta(\xi-\xi')$ becomes the Kronecker delta
$\delta_{\xi\xi'}$. We also note that the associated Legendre
functions satisfy~\cite{Balazs:1986uj, Argyres:1988ga} the equation
\begin{equation}
\Delta_{\Omega}\Big[P_{l}^{m}(\cosh\theta)e^{im\varphi}\Big]=l(l+1)P_{l}^{m}(\cosh\theta)e^{im\varphi}~,
\end{equation}
or equivalently
\begin{equation}\label{angderPe}
\Delta_{\Omega}\Big[P_{-\frac{1}{2}+i\xi}^{m}(\cosh\theta)e^{im\varphi}\Big]=-\left(\xi^{2}+\frac{1}{4}\right)
P_{-\frac{1}{2}+i\xi}^{m}(\cosh\theta)e^{im\varphi}~,
\end{equation}
where $\Delta_{\Omega}$ is the differential operator
\begin{equation}
\Delta_{\Omega}=\frac{1}{\sinh\theta}\partial_{\theta}\big(\sinh\theta\partial_{\theta}\big)+
\frac{1}{\sinh^{2}\theta}\partial_{\varphi}^{2}~.
\end{equation}

We define the functions $\Ycksi$ as
\begin{equation}
\begin{split}\label{defYcksi}
\Ycksi(\theta,\varphi)&\equiv\Bigg(\frac{2\pi}{\xi\tanh(\pi\xi)}\Bigg)^{\frac{1}{2}}\Pcksi(\cosh\theta)e^{im\varphi}\\
&=\Bigg(\frac{2\pi}{\xi\tanh(\pi\xi)}\Bigg)^{\frac{1}{2}}\frac{\Gamma(i\xi+\frac{1}{2})}{\Gamma(i\xi+m+\frac{1}{2})}
P_{-\frac{1}{2}+i\xi}^{m}(\cosh\theta)e^{im\varphi}~.
\end{split}
\end{equation}
From this definition and the equation (\ref{angderPe}) we see that
\begin{equation}\label{angderYApp}
\Delta_{\Omega}\Ycksi(\theta,\varphi)=-\bigg(\xi^{2}+\frac{1}{4}\bigg)\Ycksi(\theta,\varphi)~.
\end{equation}
The functions $\Ycksi$ form a complete set of functions on the
pseudosphere $H^{2}$. For the two-dimensional manifold
$\Sigma=H^{2}/\Gamma$, which is a compact manifold of genus
$\gen=2$, they form a complete set of functions, $\xi$ takes
discrete real values and they must satisfy four periodicity
conditions, since the functions $\Pc(\cosh\theta)$ satisfy four
periodicity conditions~\cite{Balazs:1986uj}, due to the compactness
of $\Sigma$. The orthogonality condition of the $\Ycksi$ is found
from the equation (\ref{orthoPcksi}) to be
\begin{equation}\label{orthoApp}
\int_{0}^{\infty}d\theta\int_{0}^{2\pi}d\varphi\sinh\theta\Ycksi(\theta,\varphi)
\Big(\Ycksip(\theta,\varphi)\Big)^{*}=\delta_{\xi\xi'}\delta_{mm'}~.
\end{equation}

\section{Appendix: Dimensional Reduction for the Scalar Hair of the MTZ Black Hole
in the Conformal Frame}


We are going to perform the dimensional reduction procedure for
the action (\ref{Iconfframe}). We consider the region near the
event horizon $r_{+}$, which is the only event horizon for
non-negative masses and the outermost event horizon for negative
masses. In this region, as we have previously seen, we have
\begin{equation}\label{rintortcApp}
r(r_\ast)\approx A e^{2\kappa r_\ast}+r_{+}~,
\end{equation}
and
\begin{equation}\label{fintortcApp}
f(r(r_\ast))\approx 2\kappa Ae^{2\kappa r_\ast}~.
\end{equation}
Hence, the limit $r\rightarrow r_{+}$ is equivalent to
$r_\ast\rightarrow-\infty$ in tortoise coordinates and
$f(r(r_\ast))$ is a suppression factor near the horizon. After
transforming to tortoise coordinates the action (\ref{Iconfframe})
takes the form
\begin{equation}
\begin{split}\label{IconftorApp}
I[\hat{g}_{\mu\nu},\Psi]_{\ast}=\int
dtdr_{\ast}d\Omega_{\Sigma}r^{2}(r_{\ast})f(r(r_{\ast}))&\biggl[\frac{\hat{R}(r(r_{\ast}))+6l^{-2}}{16\pi
G}-\frac{1}{2}\hat{g}^{\mu\nu}\partial_{\mu}\Psi(r(r_{\ast}))\partial_{\nu}\Psi(r(r_{\ast}))\\
&-\frac{1}{12}\hat{R}(r(r_{\ast}))\Psi^{2}(r(r_{\ast}))-\frac{2\pi
G}{3l^{2}}\Psi^{4}(r(r_{\ast}))\biggr]~,
\end{split}
\end{equation}
where $d\Omega_{\Sigma}=\sinh\theta d\theta d\varphi$. Now, we
examine the behaviour of each term of this action, using the
equations (\ref{rintortcApp}) and (\ref{fintortcApp}). The scalar
field in tortoise coordinates near the event horizon is
\begin{equation}
\Psi(r(r_{\ast}))=\sqrt{\frac{3}{4\pi
G}}\frac{G\mu}{r((r_{\ast}))+G\mu}\;\longrightarrow\;\sqrt{\frac{3}{4\pi
G}}\frac{G\mu}{r_{+}+G\mu}~,
\end{equation}
and for the rest terms of the action (\ref{IconftorApp}) in the
region near the event horizon we get
\begin{equation}\label{1termI}
\int dr_{\ast}r^{2}(r_{\ast})f(r(r_{\ast}))\biggl(-\frac{2\pi
G}{3l^{2}}\Psi^{4}(r(r_{\ast}))\biggr)\:\longrightarrow{0}~,
\end{equation}
\begin{equation}\label{2termI}
\int dr_{\ast}r^{2}(r_{\ast})f(r(r_{\ast}))\frac{6l^{-2}}{16\pi
G}\:\longrightarrow{0}~.
\end{equation}
Substituting $\hat{R}$ from the equation (\ref{RMTZ}) to the third
term of the action (\ref{IconftorApp}) we get
\begin{equation}\label{3termI}
\int
dr_{\ast}f(r(r_{\ast}))l^{-2}\Psi^{2}(r(r_{\ast}))\:\longrightarrow{0}~.
\end{equation}
Similarly, we find
\begin{equation}\label{4termI}
\int
dr_{\ast}r^{2}(r_{\ast})f(r(r_{\ast}))\frac{\hat{R}(r(r_{\ast}))}{16\pi
G}\:\longrightarrow{0}~.
\end{equation}
The remaining term of the action to be examined is
\begin{equation}
-\frac{1}{2}\int
dtdr_{\ast}d\Omega_{\Sigma}r^{2}(r_{\ast})f(r(r_{\ast}))
\hat{g}^{\mu\nu}\partial_{\mu}\Psi(r(r_{\ast}))\partial_{\nu}\Psi(r(r_{\ast}))~,
\end{equation}
and originates from the part of the action (\ref{Iconfframe})
which is
\begin{equation}
\begin{split}
-\frac{1}{2}\int
d^{4}x\sqrt{-\hat{g}}\hat{g}^{\mu\nu}\partial_{\mu}\Psi(r)\partial_{\nu}\Psi(r)&=
-\frac{1}{2}\int
d^{4}x\sqrt{-\hat{g}}\hat{g}^{rr}\partial_{r}\Psi(r)\partial_{r}\Psi(r)\\
&=-\frac{1}{2}\int
d^{4}x\sqrt{-\hat{g}}f(r)\frac{3G\mu^{2}}{4\pi}\frac{1}{(r+G\mu)^4}~.
\end{split}
\end{equation}
Therefore, in tortoise coordinates and always near the horizon, from
the last equation, we get
\begin{equation}
-\frac{3G\mu^{2}}{8\pi}\int
dtdr_{\ast}d\Omega_{\Sigma}r^{2}(r_{\ast})f^{2}(r(r_{\ast}))\frac{1}{(r(r_{\ast})+G\mu)^{4}}\:\longrightarrow{0}~,
\end{equation}
that is
\begin{equation}\label{5termI}
-\frac{1}{2}\int
dtdr_{\ast}d\Omega_{\Sigma}r^{2}(r_{\ast})f(r(r_{\ast}))
\hat{g}^{\mu\nu}\partial_{\mu}\Psi(r(r_{\ast}))\partial_{\nu}\Psi(r(r_{\ast}))\:\longrightarrow{0}~.
\end{equation}
Adding the expressions (\ref{1termI}), (\ref{2termI}),
(\ref{3termI}), (\ref{4termI}) and (\ref{5termI}) we find that the
action (\ref{IconftorApp}) in tortoise coordinates and in the region
near the event horizon vanishes. Thus, in the vicinity of the event
horizon the action of the conformally coupled scalar field of the
MTZ black hole is
\begin{equation}
I[\hat{g}_{\mu\nu},\Psi]=0~.
\end{equation}


\end{document}